\documentclass[sigconf]{acmart}

\usepackage{subfigure}

\setcopyright{acmlicensed}
\copyrightyear{2021}
\acmYear{2021}

\acmConference[OzCHI '21]{33rd Australian Conference on Human-Computer Interaction}{November 30-December 2, 2021}{Melbourne, VIC, Australia}
\acmBooktitle{33rd Australian Conference on Human-Computer Interaction (OzCHI '21), November 30--December 2, 2021, Melbourne, VIC, Australia}
\acmPrice{15.00}
\acmDOI{10.1145/3520495.3520525}
\acmISBN{978-1-4503-9598-4/21/11}

\begin{document}

\title{Designing Trustworthy User Interfaces}

\author{Valentin Zieglmeier}
\affiliation{%
	\institution{Technical University of Munich}
	\city{Munich}
	\country{Germany}}
\email{valentin.zieglmeier@tum.de}
\orcid{0000-0002-3770-0321}

\author{Antonia Maria Lehene}
\affiliation{%
	\institution{Technical University of Munich}
	\city{Munich}
	\country{Germany}}
\email{antonia.lehene@tum.de}
\orcid{0000-0002-2064-5602}

\begin{abstract}
	Interface design can directly influence trustworthiness of a software. Thereby, it affects users' intention to use a tool.
	Previous research on user trust has not comprehensively addressed user interface design, though. We lack an understanding of what makes interfaces trustworthy (1), as well as actionable measures to improve trustworthiness (2).

	We contribute to this by addressing both gaps. Based on a systematic literature review, we give a thorough overview over the theory on user trust and provide a taxonomy of factors influencing user interface trustworthiness. Then, we derive concrete measures to address these factors in interface design. We use the results to create a proof of concept interface. In a preliminary evaluation, we compare a variant designed to elicit trust with one designed to reduce it.
	Our results show that the measures we apply can be effective in fostering trust in users.
\end{abstract}



\maketitle

\section{Introduction}

When creating software tools with user-facing interfaces, one of the central aspects to consider is how they are designed.
As this is the component that users are directly exposed to, it can shape their impression of the tool and willingness to use it.
How trustworthy the software appears at first glance may play an important role in this.
User trust has been shown to influence users' intention to use a software~\cite{Komiak2006}.
Furthermore, central trust antecedents, such as credibility, increase intention to use as well~\cite{Ong2004}.
Previous research suggests that trust constructs may even have a higher influence on intention to use than some usability aspects~\cite{wang2006predicting}.
Beyond the actual service or underlying implementation, the design of the interface has been shown to play an important role in the trustworthiness of a software tool~\cite{Pu2007}.
Yet, to the best of our knowledge, no comprehensive overview over the relevant trust constructs and how they can be concretely addressed with the design of user interfaces exists today.

Our goal is to derive how (initial) user trust can be achieved and improved through user interface design. For this purpose, general influences on user trust for various usage contexts are summarized. Additionally, we analyze the interactions of these factors, as well as how they can be implemented in user interfaces in order to initiate user trust.
In a preliminary empirical study, we assess whether a proof of concept design developed according to our findings can in fact increase user trust as we expect it.

Therefore, this work contributes a theoretical overview over user trust formation and factors in software design, summarizes actionable design variants to improve the trustworthiness of a software tool through its user interface, and provides preliminary evaluation results on the effectiveness of the developed software design variants.

\section{Systematic literature review}

From literature reviews in the field of user experience and trust~\cite{Hassenzahl2006, AKRAM2018417, Hancock2011, Hancock2011a, Hoff2015}, we derived central terms used in research on trust in automation.
These terms were used to build queries for a systematic literature review covering Scopus and Web of Science. This meant we included works that focus on individual factors related to trust.
For each term, we built a query of the form \texttt{("user trust") AND (<term> OR <synonyms>) AND NOT ("social media" OR "blockchain")}. For the search terms and synonyms we used, refer to Table~\ref{tab:terms-and-synonyms}. We explicitly excluded works with the terms ``social media'' and ``blockchain'', as we found that these often consider users' trust in each other, rather than in the respective tool.
The search covered title, abstract, and keywords. We limited the results to English language journal articles and conference papers in the area of computer science.

\begin{table}[htbp]
	\centering
	\caption{Terms and synonyms used in the search queries.}
	\label{tab:terms-and-synonyms}
	\begin{tabular}{ l l }
		\toprule
		\textbf{Term} & \textbf{Synonyms} \\
		\midrule
		interface design & aesthetics \\
		intention to use & tool adoption \\
		perceived credibility & authenticity \\
		functionality    & reliability, accuracy \\
		usability        & user satisfaction \\
		learnability     & learning method \\
		predictability   & consistency \\
		perceived security & --- \\
		personalization  & customization \\
		preference       & --- \\
		familiarity      & previous experience \\
		feedback         & communication \\
		\bottomrule
	\end{tabular}
\end{table}

In total, 697 works were found. Through a title and abstract review, we filtered for papers related to our research. This left us with 162 remaining works. These, plus 40 works that we added through snowballing, were read and analyzed.
The following sections summarize the results from the most relevant of these works.

\section{Fundamentals: User trust}

To better understand our findings, it is important to grasp what the concept of user trust encompasses. Furthermore, it is instructive to consider how trust is built and maintained, as the trust relationship is dynamic.
User trust can be defined using different approaches. In the following, we describe various aspects to understand and classify it.

\paragraph{Affect-based and cognitive trust.} First, we can differentiate between the type of processing, specifically between affect-based and cognitive trust. Cognitive trust is based on users' perception of system reliability, relating to logical reasoning~\cite{Yuksel2017, Corritore2003}. Affect-based or emotional trust on the other hand is influenced by their perception of the system's aesthetics and subjective beauty, relating to users' feelings~\cite{Corritore2003, Komiak2006, Merritt2011}.
In general, affect-based trust seems to have greater impact on overall trust~\cite{Yuksel2017}, but may be indirectly influenced by cognitive trust levels~\cite{Komiak2006}.

\begin{figure}[htbp]
	\centering
	\includegraphics[height=14em]{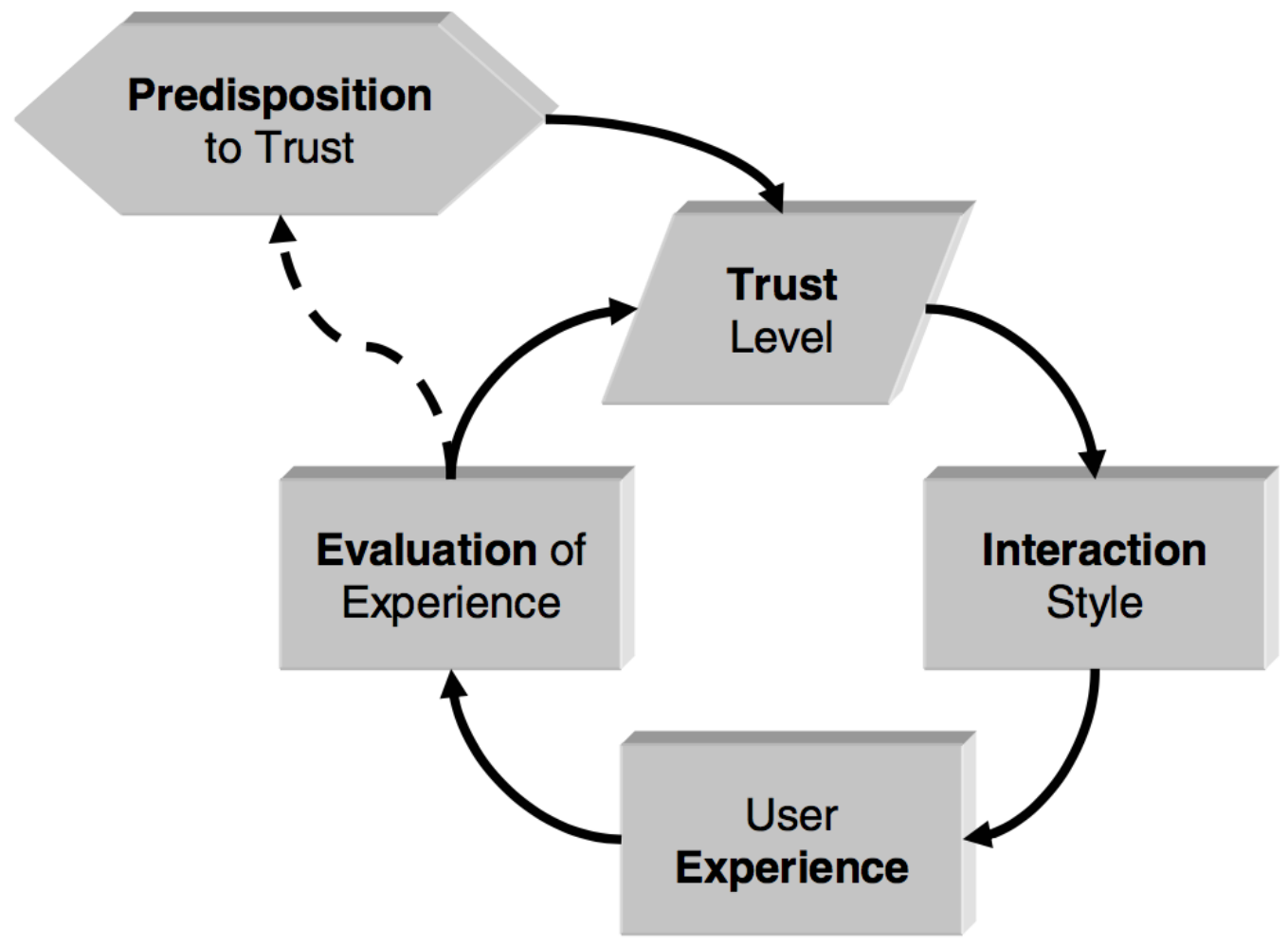}
	\caption{\citeauthor{Constantine2006}'s feedback model of trust in user-system interaction~\cite{Constantine2006}.}
	\label{fig:feedback-model-of-trust}
	\Description{A process model. At its core are four boxes, connected by arrows going in a circle from one to the next. The boxes are, in order, Trust level, Interaction style, User experience, and Evaluation of experience. Additionally, a fifth box labeled Predisposition to trust is on the left, with an arrow to Trust level, and a dashed arrow incoming from Evaluation of experience.}
\end{figure}

\paragraph{Trust life cycle.} Next, we consider the stages in the trust relationship. This can be likened to a life cycle~\cite{Constantine2006}, with initial trust transforming into long-term trust with continued use of a system~\cite{Corritore2003}.
Trust can be initially fostered, increased, decreased, lost, and regained~\cite{Corritore2003, Parasuraman1997}. In our work, we focus on initial trust, yet it is just as important to maintain continued user trust through the trust life cycle (see Figure~\ref{fig:feedback-model-of-trust}).

\paragraph{Influences on user trust.} Finally, we zoom out to recognize external influences on user trust that lie outside our control as system designers.
\emph{User criteria} are individual factors that differ between users. These encompass the predisposition or propensity to trust (see Figure~\ref{fig:feedback-model-of-trust} and, e.g., \cite{Merritt2008, Merritt2013, Constantine2006}), past experiences and familiarity with a software~\cite{Joost2004, Komiak2006}, culture~\cite{Cyr2013, Cyr2010, Chien2018}, and other individual factors such as mental state~\cite{Hoff2015}.
Influencing factors extend beyond an individual, also encompassing their \emph{environment}. While these are still somewhat individual, different users may experience them the same way. Examples are the specific task or work environment~\cite{Hoff2015}, word of mouth~\cite{Cheung2008}, other factors relating to the usage environment~\cite{Oleson2011}, or, importantly, the cultural context~\cite{Chien2014, Chien2018, Cyr2010, Faisal2017, Cyr2013}.

\section{Trustworthiness factors in interface design}
\label{sec:trust-factors}

To understand how to affect user trust through user interface design, we differentiate between trustworthiness factors, or antecedents.
We describe factors that can be influenced directly through interface design changes.
In previous works, no such overview exists.
Therefore, we summarize and combine different taxonomies and classifications of user trust.
For each factor in our list, we reference the taxonomies that include it.
We selected only those factors for which we found concrete examples in literature describing how they can be addressed through user interface design.
Those examples, comprising exemplary instantiations and empirical evaluations, are referenced as well.
The result is a robust taxonomy of trustworthiness factors in interface design (see Table~\ref{tab:trust-factors}).

\begin{table*}[htbp]
	\centering
	\caption{Trustworthiness factors identified in literature. \emph{Taxonomies} lists all found taxonomies or theoretical models that include this factor. \emph{Examples} lists exemplary instantiations or empirical evaluations of this factor.}
	\begin{tabular}{ l l l l }
		\toprule
		\textbf{Dimension} & \textbf{Factor} & \textbf{Taxonomies} & \textbf{Examples} \\
		\midrule
		Purpose     & Benevolence        & \cite{Lee2004, Constantine2006, Sutcliffe2006, sollner2011towards} & \cite{Cheung2008, Hoffman2013, sousa2014design, harwood2017internet} \\
		            & Credibility        & \cite{Fogg1999, Tseng1999} & \cite{Head, Corritore2005, Flanagin2007, Hammer2015} \\
		            & Perceived security & \cite{Kirlappos2012, Kim2007, Tarafdar2005} & \cite{Kirlappos2012, Kim2007, Chen2007, Alraja2019} \\
		\addlinespace{}%
		Process     & Integrity          & \cite{Lee2004, Sutcliffe2006, sollner2011towards, french2018trust, jian2000foundations} & \cite{Kirlappos2012, Kim2007, Cheung2008, Hoffmann2013} \\
		            & Predictability     & \cite{Lee2004, Sutcliffe2006, Constantine2006, Hancock2011a, sollner2011towards, french2018trust} & \cite{Roy2001, Hoffmann2013, Perkins2010, sousa2014design} \\
		            & Transparency       & \cite{Lee2004, Constantine2006, Hancock2011a, Hoff2015, sollner2011towards, french2018trust, kelly2003guidelines} & \cite{Hammer2015, Oduor2007, Antifakos2005} \\
		            & Familiarity        & \cite{french2018trust, kelly2003guidelines, jian2000foundations, Kirlappos2012, Du2009} & \cite{Gefen2000, Begany2015, Holzinger2011, Pirhonen2005} \\
		            & Communication      & \cite{Hancock2011a, Hoff2015, french2018trust, Mirnig2014} & \cite{Hoffmann2013, Mayeh2016, Serge2015, Antifakos2005} \\
		            &  Usability          & \cite{Lee2004, Sutcliffe2006, Hoff2015, Kirlappos2012, Kim2007, Du2009, Tarafdar2005} & \cite{Flavian2006, Roy2001, Lee2009, Lee2015} \\
		            & Personalization    & \cite{sollner2011towards, Kirlappos2012, Bernhaupt, Tarafdar2005} & \cite{Komiak2006, Li2013} \\
		\addlinespace{}%
		Performance & Competence         & \cite{Lee2004, Sutcliffe2006, Constantine2006, sollner2011towards, french2018trust, kelly2003guidelines, Kirlappos2012, Tarafdar2005} & \cite{Hoffmann2013, Komiak2006, Perkins2010, Begany2015} \\
		            & Reliability        & \cite{Parasuraman1997, Lee2004, Sutcliffe2006, Hancock2011a, Hoff2015, sollner2011towards, french2018trust, kelly2003guidelines, jian2000foundations} & \cite{Chien2018, Hammer2015, Oduor2007, Stanton2009} \\
		            & Validity           & \cite{Hancock2011a, Hoff2015, sollner2011towards, kelly2003guidelines, Pattanaphanchai2013, Tarafdar2005} & \cite{Hoffman2013, Stanton2009} \\
	  \bottomrule
	\end{tabular}
	\label{tab:trust-factors}
\end{table*}

We differentiate between trustworthiness factors by assigning them to the (perceived) purpose, process, and performance of the system~\cite{Hoff2015, Lee2004, lee1992trust}.

\subsection{Purpose dimension}

The \emph{purpose} of the system depends on its intended use~\cite{Hoff2015}. Trustworthiness factors related to this dimension reflect the impression of the designer's intentions that users get from interacting with the system~\cite{Hoffmann2014, Lee2004}.

\emph{Benevolence.} The trust relationship depends fundamentally on users' belief in benevolence from the trustee, in our case the system~\cite{Constantine2006}. A benevolent system handles user data with care and respects users actions~\cite{Constantine2006}.

\emph{Credibility.} Also referred to as honesty or sincerity, this factor refers to the perceived believability of the system~\cite{Fogg1999}. To design a credible system, its interface should be built in accordance to users' expectations and mental models~\cite{Fogg1999}.

\emph{Perceived security.} While many factors depend on user perceptions, this one is solely determined by it. Through, e.g., a more complex authentication process~\cite{Yenisey2005} or visible data security statements in the interface~\cite{Hoffmann2013}, users' sense of security can be improved.

\subsection{Process dimension}

The \emph{process} dimension describes how the system operates~\cite{Hoff2015}. These factors are defined by users' perception of how appropriate the system design is for its stated purpose~\cite{Hoffmann2014, Lee2004}.

\emph{Integrity.} Reflects users' impression of the values underlying the system design~\cite{Sutcliffe2006}, and their belief that the designers acted ethically and fulfill their promises~\cite{hasan2012modelling}. For example, certification badges or brand images can convey this~\cite{Sutcliffe2006}.

\emph{Predictability.} A fundamental facet of trustworthiness, reflected in the consistency of the behavior and design of the system~\cite{Constantine2006}.
This allows users to predict the system's future actions~\cite{Hoffmann2013}.

\emph{Transparency.} This means informing the user about the tool, specifically what it does and how it works~\cite{Hoff2015, Schnorf2014}. Nicely summarized as ``the user interface parallel to honesty in human relationships''~\cite[p.~24]{Constantine2006}.

\emph{Familiarity.} If a system is not necessarily predictable or transparent, users' familiarity with it can also help them in understanding it~\cite{french2018trust}. Even if the concrete system is new, following established patterns or designs can foster this~\cite{Holzinger2011}.

\emph{Communication.} While transparency requires the system to make information easily available and understandable, its communication reflects how it actively engages users~\cite{Hoff2015, Mirnig2014}. Explicit feedback~\cite{Antifakos2005}, with short, clear, non-intrusive messages~\cite{Mirnig2014} and a positive communication style~\cite{Mayeh2016} seem to be preferable.

\emph{Usability.} A multi-faceted construct broadly concerning the quality of the tool in enabling individuals to use it. Multiple usability factors can also elicit trust~\cite{Hoffmann2013}, specifically the ease of use~\cite{Kim2007}, ease of navigation~\cite{Roy2001}, and learnability~\cite{Alves2016}.

\emph{Personalization.} The perception of users how well the system is personalized to their needs~\cite{Komiak2006}. This can be achieved by allowing them to actively customize the tool~\cite{Sutcliffe2006}. Alternatively, it can be accomplished by learning their needs automatically and reacting to them~\cite{Komiak2006}, e.g. by providing a list of their most commonly used functions for quick access.

\subsection{Performance dimension}

The \emph{performance} of the system reflects how well the system solves its tasks~\cite{Hoff2015}. Users judge what the system does and if it can help them achieve their goals~\cite{Hoffmann2014, Lee2004}.

\emph{Competence.} The primary performance measure, indicating if a system is capable to achieve its task well. This includes not just the quality of the results, but also the time it takes to deliver them~\cite{Constantine2006}.

\emph{Reliability.} The consistency of the functions of the system~\cite{Hoff2015, Hoffmann2014}, which can also be beneficial for its predictability. A reliable system completes its tasks consistently, while a predictable system operates in ways that users expect~\cite{Hoff2015}.

\emph{Validity.} The degree to which the tasks are completed by the system as intended by the user~\cite{Hoff2015}. A low reliability will incur a lowered validity as well.

\section{Designing user interfaces to elicit trust}

The factors described above are influential when trying to foster user trust. They are, however, fairly abstract concepts. To address them with a user interface, we require concrete and actionable measures.
In the following, we describe how to systematically improve trustworthiness of an interface through exemplary measures that directly target these factors in order to elicit trust.
We do not cover the factors of the \emph{performance} dimension below, though. While they can be addressed through interface design, we found more effective measures seem to include modifications to the underlying system that are out of scope for this work.

Targeting perceived \emph{benevolence}, the system should be built to be responsive to users and convey a sense of care~\cite{Constantine2006}. For example, caching user input for repeated entry~\cite{Constantine2006} or providing advice when necessary~\cite{harwood2017internet} can communicate this.

For \emph{credibility}, the foundational work by \citeauthor{Fogg1999} serves as a guide. Regarding the interface, they define the display as well as the interaction experience as relevant aspects. Their suggestion is to match users' expectations of the system~\cite[p.~85]{Fogg1999}.
This is use case specific and could be evaluated before development through surveys or similar instruments.
Generally, choosing text-based (compared to anthropomorphic or audible) interfaces seems to increase credibility for users~\cite{Tseng2014, BURGOON2000553}. Additionally, reducing the complexity of the interface is beneficial~\cite{Tseng2014}.

The \emph{perceived security} can be improved through security assurances.
A simple, yet effective, measure is to explicitly display details on the security measures, such as that encryption is performed or that data are being verified~\cite{distler2019security}. While this does not change the actual steps performed in the code, it raises awareness in users which increases their perceived security~\cite{distler2019security}.
Furthermore, forcing users to re-login after a certain period~\cite{Yenisey2005} and informing users that unauthorized accesses are blocked~\cite{Yenisey2005, miyazaki2000internet} can improve this.
These findings can be summarized as actively making users aware of the (ostensible) security measures that are implemented.

For \emph{integrity}, the interface ideally conveys the ethics of the designers. This can be achieved by adding e.g. certification badges or brand images~\cite{Sutcliffe2006} that suggest an ethics code or value system.

The \emph{predictability} of the interface can be achieved following the guidelines set forth by \citeauthor{gram1996software}. They suggest deterministic design that maps the observable state directly to system events, with the system providing users with information about its state and the actions they can take. Furthermore, they describe completeness and consistency of information display as important~\cite[p.~296]{gram1996software}.

Increasing the \emph{transparency} that users experience can be a vital aspect.
As a basic measure, explanatory texts should make clear what functionality exists~\cite{Bigras2018}.
Especially for safety-relevant scenarios, the system should also be transparent about the risks and limitations of its functions~\cite{Kunze2019, Perkins2010, Corritore2005}.

Evoking \emph{familiarity} is of course very dependent on the users' previous experience. Still, just by following established design patterns and metaphors, users' familiarity with them can be evoked~\cite{Holzinger2011}.

Regarding \emph{communication}, short and clear notifications about the status of the system serve to inform users~\cite{Sutcliffe2006, Faisal2017}.
When designing these messages, etiquette are important~\cite{Parasuraman2004, Lee2004, Xu2014}.
These can be defined as being non-interrupting and patient~\cite{Parasuraman2004}, as well as having a positive tone~\cite{Hartmann2008}.

The \emph{usability} of a user interface is a more complex topic and its own branch of research.
Yet, from works addressing usability as an antecedent of trust specifically, we can derive some concrete suggestions.
First, ease of navigation and user guidance are beneficial~\cite{Roy2001, Sillence2004, Faisal2017}.
Similarly, consistency in design and color schemes improves usability and trustworthiness~\cite{Sutcliffe2006, Faisal2017, Cyr2010}.
For non-intuitive interfaces, learnability was found to be effective~\cite{Roy2001, Sillence2004, AtkinsonUAHCI2007}. This can mean giving users the opportunity to learn about the functions of the system and encouraging them to explore it~\cite{AtkinsonUAHCI2007}. If training is required, tools can directly embed tutorials to ease discovery~\cite{Tognazzini}.
Additionally, the ease of use and subjective appeal of the interface can be relevant~\cite{Gao2005, Vila2011, Sonderegger2014}. Beyond generally aiming to reduce the required cognitive effort, this can mean improving the reaction speed~\cite{Vila2011}, reducing clutter and animations~\cite{Gao2005}, and designing a layout with, e.g., high classical aesthetic appeal~\cite{Sonderegger2014}.
Finally, attractive as well as readable typography, covering font choice and text size, are also facets to consider~\cite{Faisal2017, Fisher2008}.

For \emph{personalization}, various measures can be effective. Allowing customization of the interface to match the user needs~\cite{Sutcliffe2006} is a sensible step, while more advanced systems might try to predict user wishes~\cite{Komiak2006}.

\section{Preliminary evaluation}

To assess the effectiveness of influencing trust through user interface design, we created a proof of concept interface for a preliminary evaluation. We developed two user interface variants: One variant aimed to follow our recommendations to elicit trust (variant A), the other explicitly disregarded our findings (variant B).
Twelve people participated in our study (ten students, two apprentices; eight female, four male). Each participant was asked to assess both variants. They had to register, then they were led to variant A. After exploring it and answering the questionnaire, they were then shown variant B. Finally, they answered the same questionnaire again to allow us to compare the results.
Their responses were recorded on a seven-point Likert scale.

\subsection{Design variants}

\begin{figure}[htbp]
	\centering
	\subfigure{
		\includegraphics[width=1\linewidth]{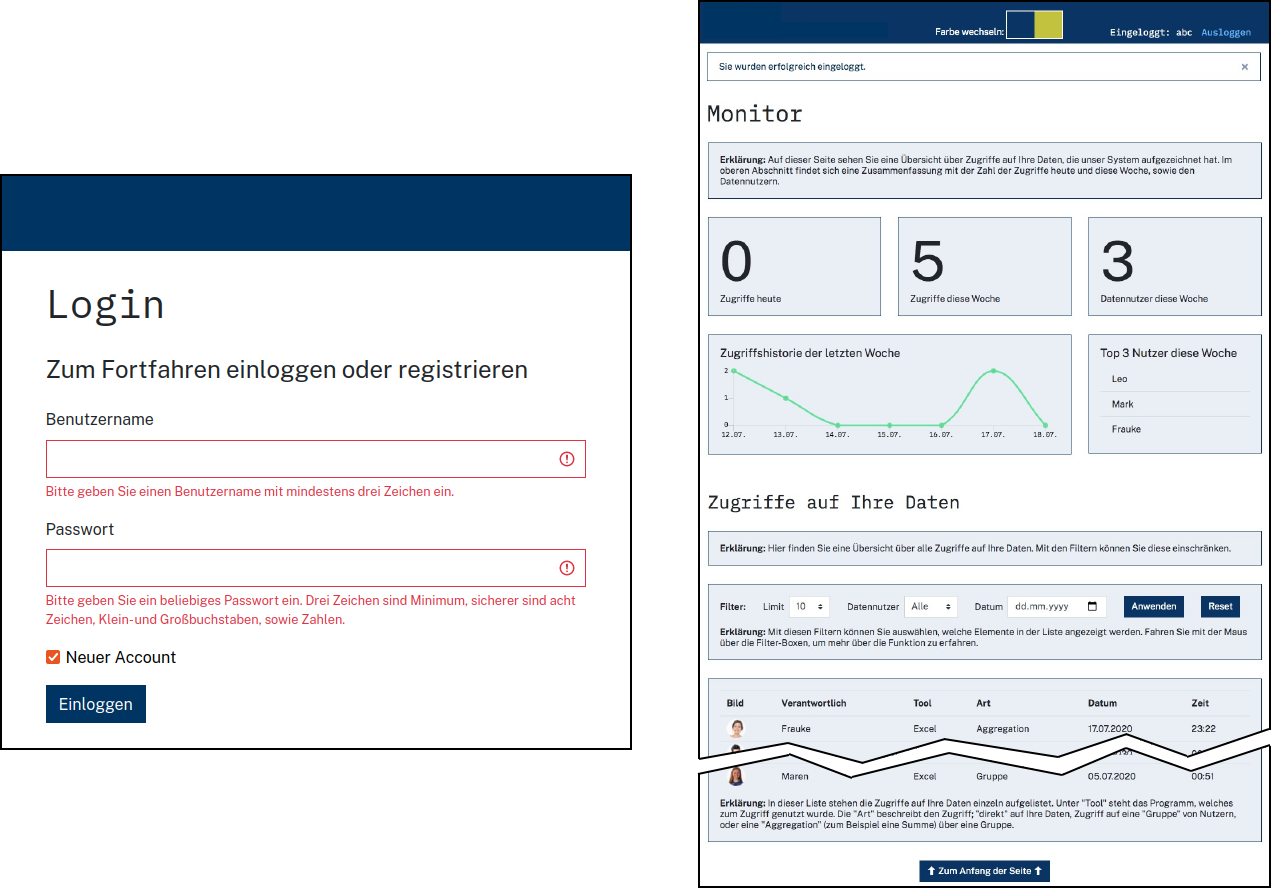}
		\label{fig:screenshot-A-combined}
	}
	\subfigure{
		\includegraphics[width=1\linewidth]{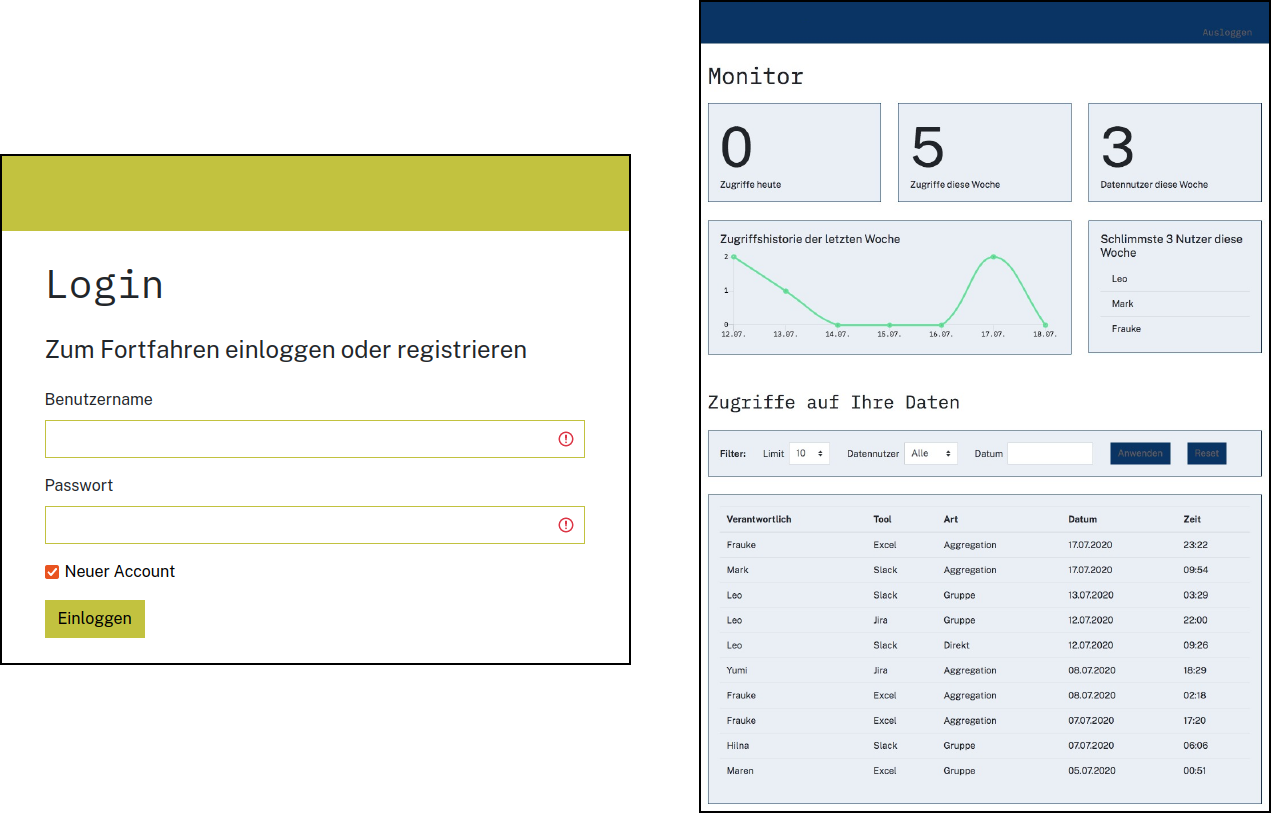}
		\label{fig:screenshot-B-combined}
	}
	\caption{Screenshots of the two interface design variants. Variant A (above) was developed to elicit trust, variant B (below) to reduce it.}
	\label{fig:screenshots}
	\Description{Four screenshots, two for each design variant. On the left are the two login screens, on the right two dashboards, showing graphs and numbers with illegible text. Their concrete style differences are described in the text of this section.}
\end{figure}

We created two interface variants for participants to explore (see Figure~\ref{fig:screenshots}). Both followed the same basic structure: They started with a login page where users had to authenticate, followed by the main page in form of a dashboard listing various data.
The dashboard was designed for the use case of logging data usages~\cite[see, e.g.,][]{angulo2015usable, zieglmeier2021trustworthy}, showing how data of the individual were accessed and by whom.

To target personalization, variant A offered a color scheme picker, while variant B did not.
For transparency and communication, a message was displayed after successful login, with a status bar showing the logged-in user. Both were missing from variant B.
A more positive communication style for variant A meant that it listed the ``best'' users, instead of the ``worst'', when showing who accessed the data most frequently.
Improving predictability, a detailed description of available functions was shown only for variant A. Also, the same color scheme was used for the login page and dashboard, whereas for variant B it switched from yellow to blue.
Finally, we targeted usability.
To improve ease of navigation, variant A had a button at the bottom to return ``back to top'' that B did not have. Additionally, the ``Log out'' button on variant A was made easier to notice with brighter text.
For ease of use and subjective appeal, the colors of text on buttons for variant B in general were made less legible compared to A, with a reduced contrast to the background color.
Furthermore, a semaphore informing users how strong the password they selected was and if it was valid was shown on variant A, but not on B.

\subsection{Questionnaire}

Participants answered ten questions related to our design changes, with their responses being recorded on a seven-point Likert scale. Lower values on the scale indicate lower levels of agreement.
We used questions developed and validated in previous works to evaluate trust.

Covering the induced familiarity and predictability, we asked:

\begin{itemize}
	\item FP: I am familiar with how the system works~\cite[from][]{Komiak2006}.
\end{itemize}

For transparency, learnability, and communication, we asked:

\begin{itemize}
	\item TLC: I find the system easy to learn to use~\cite[from][]{Faisal2017}.
\end{itemize}

Regarding their perceived security, we asked:

\begin{itemize}
	\item PS: I think the authentication is very secure, that is, it protects me against attacks~\cite[from][]{Zimmermann2020}.
\end{itemize}

Addressing usability directly, we asked:
\begin{itemize}
	\item U1: I find the system easy to use~\cite[from][]{Corritore2005}.
	\item U2: I can find easily what I am looking for on the interface~\cite[from][]{Roy2001}.
\end{itemize}

To assess their faith in the system generally, we asked:

\begin{itemize}
	\item F1: I believe advice from the system even when I don’t know for certain that it is correct~\cite[from][]{Madsen2000}.
	\item F2: When I am uncertain about a decision I believe the system rather than myself~\cite[from][]{Madsen2000}.
	\item F3: If I am not sure about a decision, I have faith that the system will provide the best solution~\cite[from][]{Madsen2000}.
\end{itemize}

In addition, we asked participants about their trust and the perceived trustworthiness of the system directly:

\begin{itemize}
	\item T1: I trust the system~\cite[from][]{Corritore2005, Merritt2011}.
	\item T2: I believe the system to be trustworthy~\cite[from][]{Everard2006, Corritore2005, Hammer2015}.
\end{itemize}

\subsection{Results}

We analyze the results by comparing the median value of all responses for variant A (eliciting trust) to variant B.
We use the median instead of averaging the response values, as that can be problematic for ordinal data such as Likert responses. It prevents outliers from affecting the result while still allowing us to summarize the responses in a single value.
For visualization, we use hat graphs~\cite{witt2019introducing}.

\begin{figure}[htbp]
	\centering
	\subfigure[Questions targeting trustworthiness factors.]{
		\includegraphics[height=16.5em]{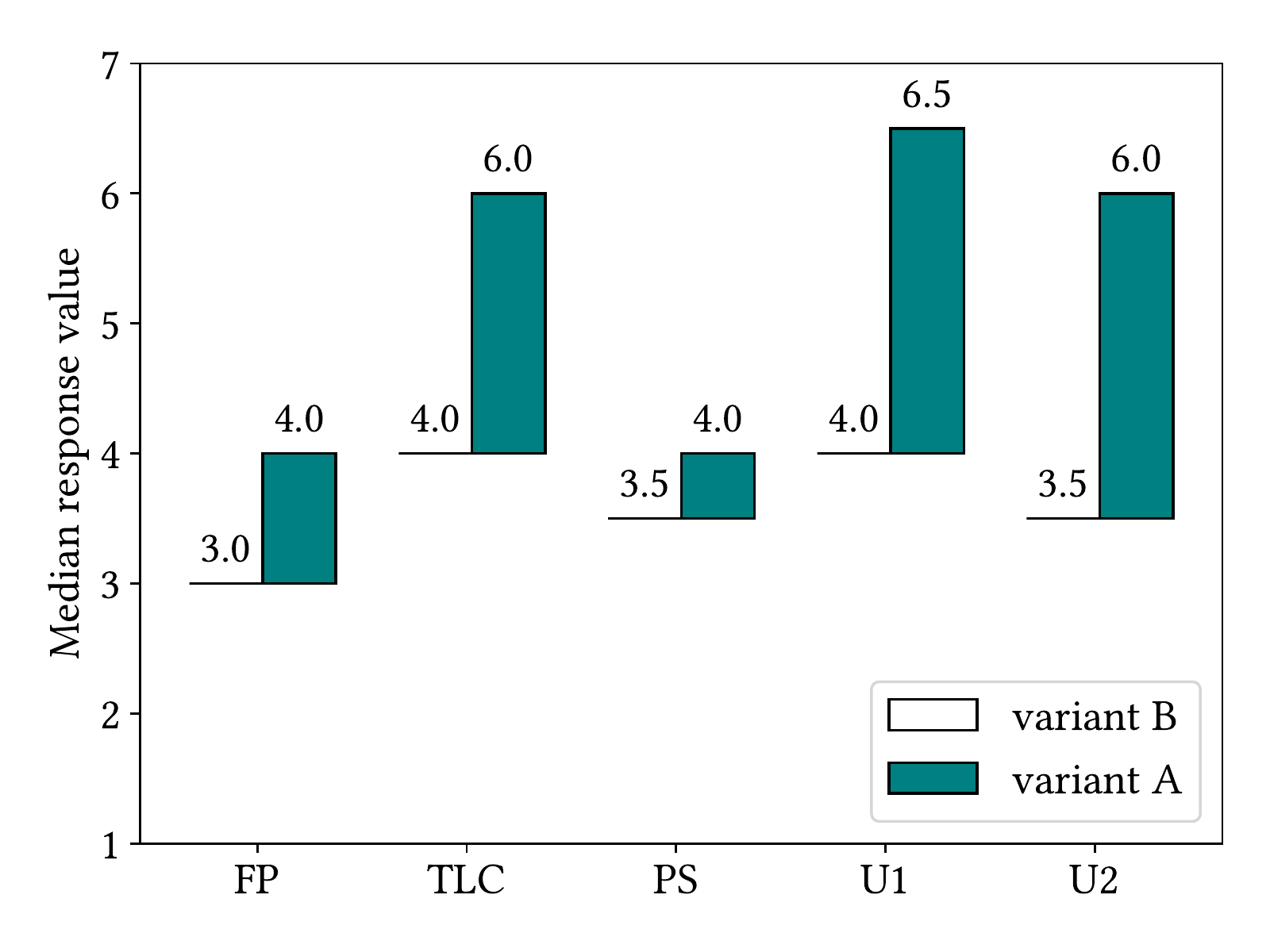}
		\label{fig:plot-trustworthiness-factors}
	}
	\subfigure[Questions targeting faith and trust broadly.]{
		\includegraphics[height=16.5em]{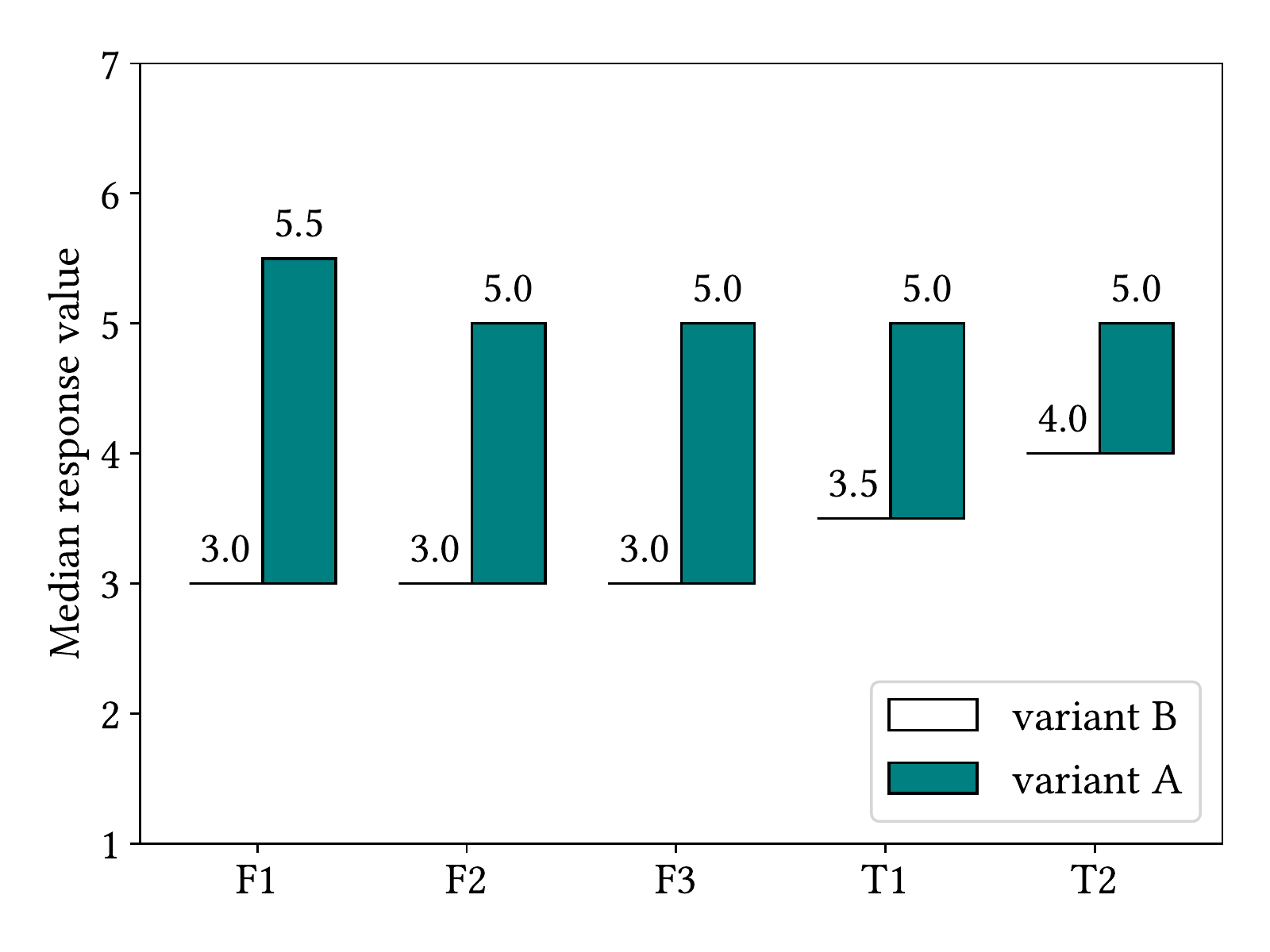}
		\label{fig:plot-faith-trust}
	}
	\caption{Questionnaire results ($n=12$). The median of all responses for variant A (trust-eliciting) and variant B is compared.}
	\label{fig:plots-results}
	\Description{Two plots with so-called hat graphs, which are similar to bar graphs. Each plot has five hats that show two values each. The values are, in the order variant A, then variant B: FP 3.0 to 4.0, TLC 4.0 to 6.0, PS 3.5 to 4.0, U1 4.0 to 6.5, U2 3.5 to 6.0. For plot 2, F1 3.0 to 5.5, F2 3.0 to 5.0, F3 3.0 to 5.0, T1 3.5 to 5.0, T2 4.0 to 5.0.}
\end{figure}

For most trustworthiness factors, we find a clear increase in the median level of agreement (see Figure~\ref{fig:plot-trustworthiness-factors}). On average, the median increased by $1.7$ points, with the lowest increase ($0.5$) for question PS and the highest ($2.5$) for questions U1 and U2.
This suggests that our changes had the greatest impact on usability.
Considering questions FP and TLC, we find an interesting difference: participants seemingly did not understand how the system worked when they first used it, but may have found they could learn to do so.
The low difference for question PS seems plausible, as our modifications did not explicitly target perceived security and both variants used the same authentication process.

Considering the overarching goal of trust and the related construct of faith, we also find noticeable increases in the median level of agreement (see Figure~\ref{fig:plot-faith-trust}). The average increase is $1.8$ points, with the lowest ($1$) for question T2 and the highest ($2.5$) for question F1.
For variant A, we can see that the median is stable at $5$, except for F1 with $5.5$. This means that participants did not overwhelmingly agree with the questions, but showed a clear tendency towards agreement. The lower increase for question T2 is not due to variant A being less trustworthy, but due to variant B seemingly also eliciting some trust in participants.
In all, this suggests that our changes had the impact we aimed for.

\section{Discussion}

At first glance, aiming to elicit trust in users solely through interface design may seem counterintuitive. Surely, confidence in a system should arise from actual and verifiable properties of the system. Yet at the same time, it seems clear that laypeople will not be able to verify such system properties in many cases. Therefore, their trust is required even then, e.g. in a third party auditor or the developers.
Instead, we believe that trust in software is important to consider on its own. Based on our findings, initial trust in a tool can be fundamentally influenced by user interface design. That means that deliberate design is beneficial and in some cases necessary to elicit trust in the developed tool.

The results from our preliminary evaluation seem to confirm our hypothesis. With very few adjustments to a proof of concept interface compared to our control, we found a noticeable increase in perceived trustworthiness by users.

Our evaluation is limited, though. Only twelve participants were part of the study, and they were shown both interface variants.
That means that we can neither exclude the influence of individual characteristics, nor an exaggerated effect by participants being able to compare the variants.

For future work, we therefore suggest expanding on these ideas and conducting a more robust and in-depth evaluation.
In addition to expanding the set of participants, assessing their individual predisposition to trust before the evaluation may help in understanding some differences in the responses.
Finally, of course, we hope our work will serve to guide and support the design of trustworthy user interfaces.

\begin{acks}
This work was supported by the German Federal Ministry of Education and Research (BMBF) under grant no. 5091121.
\end{acks}

\bibliographystyle{ACM-Reference-Format}
\bibliography{references}

\end{document}